\documentclass[apj]{emulateapj}

\slugcomment{\it ApJ Accepted--- \today}
\shorttitle{DENSE GAS IN THE CENTRAL KPC OF NGC 6946}
\shortauthors{LEVINE ET AL.}
\usepackage{natbib}

\begin{document}

\title{The Dense Gas in the Central Kiloparsec of NGC 6946}
\author{E.S. Levine, Tamara T. Helfer\altaffilmark{1}, R. Meijerink, and Leo Blitz}
\affil{Department of Astronomy, University of California at Berkeley, Berkeley, CA 94720}
\email{elevine@astron.berkeley.edu}
\altaffiltext{1}{Current address: Lawrence Livermore National Laboratory, Livermore, CA 94551}

\begin{abstract}
We present observations of the HCN and HCO$^+$ J=1-0 transitions in the center of the nearby spiral galaxy NGC 6946 made with the BIMA and CARMA interferometers. Using the BIMA SONG CO map, we investigate the change in the $I_{\rm HCN}/I_{\rm CO}$  and $I_{\rm HCO+}/I_{\rm CO}$ integrated intensity ratios as a function of radius in the central kiloparsec of the galaxy, and find that they are strongly concentrated at the  center.
We use the 2MASS $K_S$ band image to find the stellar surface density, and then construct a map of the hydrostatic midplane pressure. 
We apply a PDR model to the observed $I_{\rm HCN}/I_{\rm HCO+}$ integrated intensity ratio to calculate the number density of molecular hydrogen in the dense gas tracer emitting region, and find that it is roughly constant at $10^5$ cm$^{-3}$ across our map. We explore two hypotheses for the distribution of the dense gas. If the HCN and HCO$^+$ emission comes from self-gravitating density peaks inside of a less dense gas distribution, there is a linear proportionality between the internal velocity dispersion of the dense gas and the size of the density peak. Alternatively, the HCN and HCO$^+$ emission could come from dense gas homogeneously distributed throughout the center and bound by ambient pressure, similar to what is observed toward the center of the Milky Way. We find both of these hypotheses to be plausible. 
We fit the relationships between $I_{\rm HCN}$, $I_{\rm HCO+}$, and $I_{\rm CO}$.
Correlations between the hydrostatic midplane pressure and $I_{\rm HCN}$ and $I_{\rm HCO+}$ are demonstrated, and power law fits are provided. We confirm the validity of a relation found by \citet{BR2006} between pressure and the molecular to atomic gas ratio in the high hydrostatic midplane pressure regime ($10^6$-$10^8$ cm$^{-3}$ K). 
\end{abstract}


\section{Introduction}
How is molecular gas organized in  galaxies? In the disk of the Milky Way, molecular gas as traced by CO is concentrated in giant molecular clouds (GMCs) with a characteristic size of 50 pc \citep{B1987}. These clouds have masses around 10$^4$--10$^6$ $M_\odot$ and contain 80\% of the molecular mass \citep{B1993}. 
The hydrostatic pressure from the self-gravity of a GMC this size is $\sim10^5$ cm$^{-3}$ K \citep{B1993}, which is an order of magnitude larger than the mean pressure of the local ISM \citep{B1987a}; this leads to the conclusion that GMCs in the Milky Way disk are self-gravitating.
GMCs in M31 \citep{R2007} and M33 \citep{RB2004,RKMW2007} have similar properties to those observed in the Milky Way.

The densest parts of GMCs are traced by molecules with larger dipole moments than CO, such as HCN, HCO$^+$, and CS. These dense gas tracers need a molecular hydrogen density near $10^5$ cm$^{-3}$ to be excited collisionally, 100 times greater than CO. In the disk, emission from these molecules  is limited to the star-forming cores of GMCs \citep{PJE1992,WE2003,HB1997b}.

However, the molecular gas composition    of the central kiloparsec of a spiral galaxy is very different from that of  its disk. In the Milky Way, widespread CO emission in the inner kpc makes it  difficult to differentiate between a distribution of discrete clouds or one in which the molecular gas is organized homogeneously \citep{DUCD1987}.
The emission from  HCN and CS within $\sim$250 pc of the Galactic center is  distributed similarly to the CO \citep{JHPB1996,BSWH1987}, suggesting that high  density regions are common, at least compared to the average ISM  density of $\sim$1 cm$^{-3}$  in the solar neighborhood. 
In other galaxies, dense gas tracers are more concentrated than CO in the central kpc  and have lower intensities elsewhere \citep{GS2004,HB1997a}, consistent with observations of the Milky Way \citep{HB1997b}.
Furthermore, the centers of spiral galaxies tend to have high  densities and pressures;  bulges  have hydrostatic midplane pressures 2-3 orders of magnitude higher than those of their disks \citep{SB1992}.  \citet{BR2006} (hereafter BR06) found a simple relationship in disk galaxies between  the hydrostatic midplane pressure and the ratio of molecular to atomic gas that holds in disks and bulges.

In this paper, we use new  HCN and HCO$^+$ observations of the central region of NGC 6946 to study  molecular gas organization in this nearby spiral galaxy. 
Optically, this galaxy is classified as SAB(rs)cd \citep{D1963}, it has a moderate starburst in its nucleus \citep{TH1983}, and it is at a distance of 5.5 Mpc \citep{T1988}. At this distance, 1'' corresponds to 27 pc. Observations of NGC 6946 in the R, I, and K bands reveal a small bulge (radius 15'') within a larger  bar (radius 63'')  \citep{DTVR2003}. 
Recent studies revealed the rich details of its nuclear structure; \citet{SBEL2006} used high resolution images of CO to map the molecular gas spiral in the inner 10'', likely caused by an inner stellar bar 15'' in length (see also NIR images in \citealt{ECS1998} and \citealt{KDSB2003}). A further study used HCN emission in the central 2'' as a proxy for star formation in young clusters still embedded in their parent clouds, and combined these data with observations of HII regions to map the total amount of star formation \citep{SBED2007}. They concluded the nuclear star formation is being fed by an accumulation of dense gas driven by the inner stellar bar.

Using  HCN and HCO$^+$ observations in conjunction with CO and near-infrared data, we investigate the distribution of dense gas in the central kiloparsec of NGC 6946.
After reducing our observations, we calculate  HCN/CO and HCO$^+$/CO integrated intensity ratios, and we estimate the pressure using two different methods.
We then describe two hypotheses for the distribution of the dense molecular gas, and explore the relationship between dense gas tracers  and hydrostatic midplane pressure.

\section{Observations and Data Reduction}\label{sec:obs}

\subsection{New Observations}
\subsubsection{BIMA HCN and HCO$^+$ Observations}
We observed the HCN $J=1-0$ transition at $\nu=88.631847$ GHz simultaneously with the HCO$^+$ $J=1-0$ transition at $\nu=89.188518$ GHz using the 10-element Berkeley-Illinois-Maryland Association (BIMA) interferometer \citep{BIMA} at Hat Creek, CA. The observations consisted of five short tracks during September and October 2002, totaling 20 hours of observations in the C and D configurations. Baselines ranged from 1.9 to 26 k$\lambda$. The pointing and phase center of the observations was $\alpha$(J2000)=$20^h34^m52\fs33$, $\delta$(J2000)=$+60^\circ09'14\farcs2$, and we used a seven-point hexagonal mosaic with a Nyquist spacing of 57''.
The correlator was centered near 52 km s$^{-1}$, the LSR velocity of NGC 6946, and configured to achieve a spectral resolution of 5.3 km $s^{-1}$ over a velocity range of $-120\le v\le210$ km s$^{-1}$ for both the HCN and HCO$^+$ observations,  which extends beyond the detected velocity range of the CO emission ($\sim-50\le v\le 180$ km s$^{-1}$; \citealt{BIMASONG2}).

\subsubsection{CARMA HCN Observations}
We observed the HCN $J=1-0$ transition using the Combined Array for Research in Millimeter-wave Astronomy (CARMA)\footnote{Support for CARMA construction was derived from the states of California, Illinois, and Maryland, the Gordon and Betty Moore Foundation, the Eileen and Kenneth Norris Foundation, the Caltech Associates, and the National Science Foundation. Ongoing CARMA development and operations are supported by the National Science Foundation under a cooperative agreement, and by the CARMA partner universities.}  at Cedar Flat, CA, for 11 hours over three tracks in December 2006 and March 2007. CARMA is the merger of nine 6.1 m BIMA telescopes with six 10.4 m Owens Valley Radio Observatory dishes. In December, the array was in the first light configuration, which included baselines as short as 3.3 k$\lambda$ and as long as 31 k$\lambda$. In March, the telescopes were in C array, with baselines ranging from 8 k$\lambda$ to 101 k$\lambda$.  
We configured the correlator with two overlapping 62 MHz windows and a resolution of 0.977 MHz ($\sim$3 km s$^{-1}$). This setup corresponded to a velocity coverage of $-140\le v\le230$ km s$^{-1}$. We used a seven-point hexagonal mosaic with a 30'' Nyquist spacing of the 10.4 m primary beams.
We observed Uranus and 3C279 to calibrate the flux scale and the frequency response of the passband. Approximately every 25 minutes throughout each track, we observed a nearby quasar to calibrate the time dependence of the phase.

\subsubsection{Interferometric Data Reduction}
\begin{figure}
\includegraphics[angle=90,scale=.5]{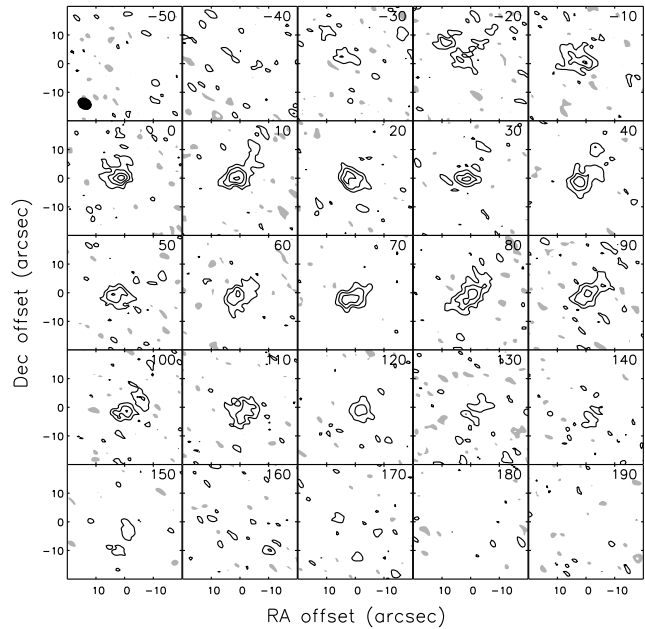}
\caption{\label{fig:channel} HCN channel map from combined BIMA and CARMA observations. The velocity
is marked in the upper right corner of each plot in km s$^{-1}$. Contour levels are drawn at 2,4,6,8\ldots times the rms channel noise. A filled grey contour is marked at -2 times the rms channel noise. The beam is shown in the lower left hand corner of the -50 km s$^{-1}$ channel.
}
\end{figure}

\begin{figure}
\includegraphics[angle=90,scale=.5]{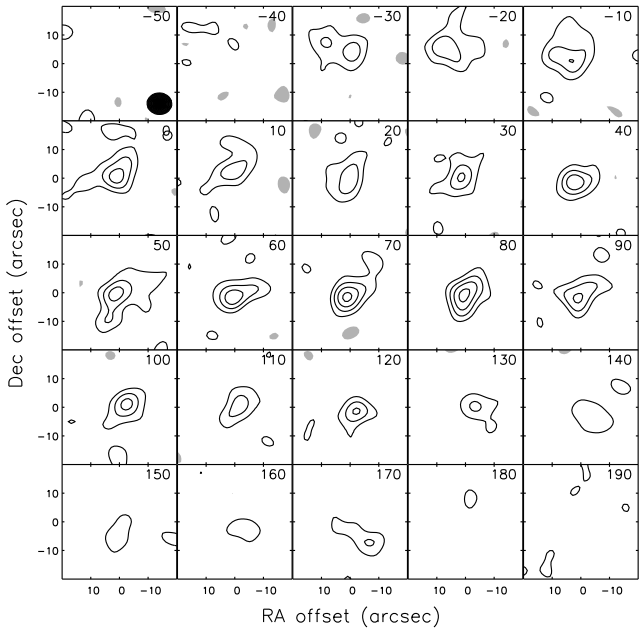}
\caption{\label{fig:hcochannel} HCO$^+$ channel map from BIMA observations. The velocity
is marked in the upper right corner of each plot in km s$^{-1}$. Contour levels are drawn at 2,4,6,8\ldots times the rms channel noise. A filled grey contour is marked at -2 times the rms channel noise. The beam is shown in the lower right hand corner of the -50 km s$^{-1}$ channel.
}
\end{figure}

We reduced the HCN and HCO$^+$ data using the MIRIAD package \citep{STW1995}. We combined the calibrated {\it uv}-visibilities and then made robustly weighted \citep{B1995} channel maps with velocity bins of width 10 km s$^{-1}$. We deconvolved the image cube using an SDI CLEAN algorithm \citep{SDI1984}; restored channel maps from this procedure are shown in Figs.~\ref{fig:channel} and \ref{fig:hcochannel}. The synthesized beam for the HCN map was 4.7'' by 3.7'' with a position angle of 61$^\circ$; for the HCO$^+$ map the beam was  8.6'' by  7.5'' with a position angle of  87$^\circ$.
We calculated the noise level by taking the average of the rms variation measured in the central arcminute box over several line-free channels on either side of the HCN or HCO$^+$ emission. The rms per 10 km s$^{-1}$ channel was  0.012 Jy/beam for the HCN map and 0.020 Jy/beam for HCO$^+$ map. Based on our flux calibration and comparison of the fluxes from individual tracks, we assigned a 1$\sigma$ error in the flux calibration of $\pm20$\%.


\subsection{CO Observations}
Observations of the $J=1-0$ line of CO at $\nu=115.2712$ GHz in NGC 6946 were taken from the BIMA Survey of Nearby Galaxies (SONG) \citep{BIMASONG1,BIMASONG2}. These observations have a resolution of 1.56 MHz ($\sim$4 km s$^{-1}$) over a total bandwidth of 368 MHz (960 km s$^{-1}$). The interferometer data was combined with observations from the NRAO 12m telescope at Kitt Peak, AZ.\footnote{The National Radio Astronomy Observatory is operated by the Associated Universities, Inc., under cooperative agreement with the National Science Foundation.} The synthesized beam was  6.0'' by  4.9'' with a position angle of 14$^\circ$. The 1$\sigma$ flux calibration error was $\pm15$\%.

\subsection{Integrated Intensity Maps}\label{sec:intensity}
\begin{figure}
\includegraphics[angle=90,scale=.5]{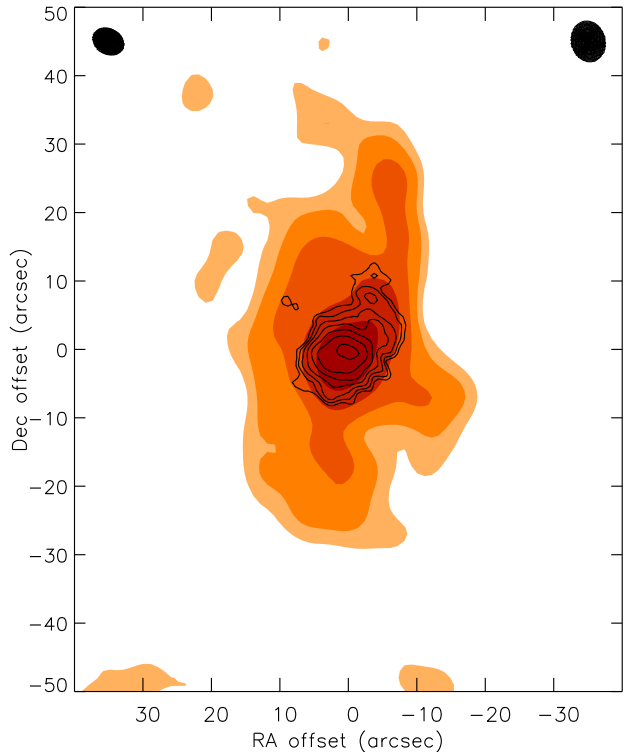}
\caption{\label{fig:intmap} HCN integrated intensity map. Red/orange colors are CO, black contour lines are HCN. Contours are logarithmically spaced in both CO and HCN, such that a contour represents 2,2$\sqrt{2}$,4,4$\sqrt{2}$,8\ldots times 0.636 Jy/beam km s$^{-1}$ for HCN and 2,4,8,16\ldots times 3.01 Jy/beam km s$^{-1}$ for CO. These factors are the $Q_X$ noise levels for each species for a point where all the channels on the line are included (see \S \ref{sec:ratio}). The HCN beam is shown in the upper left hand corner, and the CO beam is in the upper right.
}
\end{figure}

\begin{figure}
\includegraphics[angle=90,scale=.5]{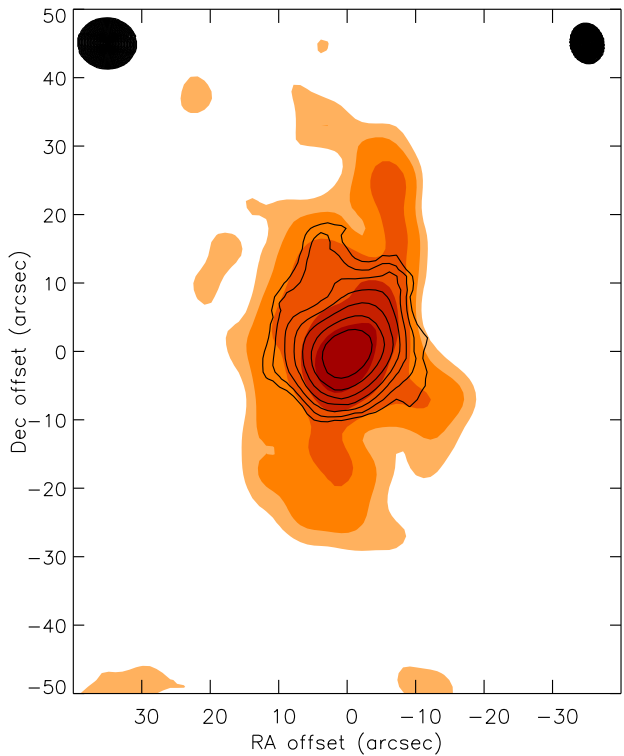}
\caption{\label{fig:hcointmap} HCO$^+$ integrated intensity map. Red/orange colors are CO, black contour lines are HCO$^+$. Contours are logarithmically spaced in both CO and HCO$^+$, such that a contour represents 2,2$\sqrt{2}$,4,4$\sqrt{2}$,8\ldots times 0.982 Jy/beam km s$^{-1}$ for HCO$^+$ and 2,4,8,16\ldots times 3.01 Jy/beam km s$^{-1}$ for CO. The HCO$^+$ beam is shown in the upper left hand corner, and the CO beam is in the upper right.}
\end{figure}

To construct integrated intensity maps of the dense gas tracers while adding in as little noise as possible, we masked out the emission in regions where the signal is below the noise threshold.  
We started with the channel maps $F_{X}(\alpha,\delta,v)$ where $X$ denotes the species and $v$ the velocity channel. We constructed independent masks for HCN and HCO$^+$ by first creating blurred maps through convolution with a 20'' Gaussian, and then averaging the rms noise in several channels off the line to calculate $\sigma_{X,{\rm blur}}$. For each of these two species  we defined a mask:
\begin{eqnarray}
 C_X(\alpha,\delta,v)&=&1~{\rm where}~F_{X}>3\sigma_{X,{\rm blur}}\nonumber\\
 &=&0~{\rm elsewhere}.
\end{eqnarray} 
 We then made integrated intensity maps: 
 \begin{equation}\label{eqn:intens}
  I_X(\alpha,\delta)=\sum_v F_{X}C_Xdv 
 \end{equation} 
  where $dv$ is the channel velocity width. This masking technique preserves noise statistics, but introduces a bias in favor of low-level diffuse emission distributed similarly to the bright emission  and against low-level compact emission distributed differently than the brightest emission (see \citet{BIMASONG1} and \citet{BIMASONG2} for more thorough discussions).
Fig.~\ref{fig:intmap} shows $I_{\rm HCN}$ contours plotted over $I_{\rm CO}$ color, and Fig.~\ref{fig:hcointmap} plots $I_{\rm HCO+}$ contours over $I_{\rm CO}$ color. For these figures, no mask was applied to the CO data.

\subsection{2MASS Image Processing}\label{sec:2mass}
We obtained 
$K_S$ (2.2 $\mu$m) band images of NGC 6946 from the 2MASS Large Galaxy Atlas \citep{JCCSH2003}. Galaxies in this survey have a typical angular resolution of 3'', with a 1'' pixel grid. First, we applied a median filter to the image to remove foreground stellar contamination. Next, we corrected the image for dust extinction from the Milky Way ($A_K$=0.125) \citep{SFD1998} as given in  NED\footnote{The NASA/IPAC Extragalactic Database (NED) is operated by the Jet Propulsion Laboratory, California Institute of Technology, under contract with the National Aeronautics and Space Administration.}. Dust in NGC 6946 will also affect the observed $K_S$ band image, but dust extinction at these wavelengths is generally small. 
Indeed, \citet{R2000} calculated a dust optical depth map for NGC 6946 by using optical and near-infrared images as inputs to a radiative transfer model. Though his model did not converge in the inner few arcseconds, he found V band extinctions peaking near 5 magnitudes just outside that region. The dust extinction varied strongly as a function of distance from the galactic center. Using the relation between the magnitude of dust absorption in $V$ and $K$ for the Milky Way \citep{SFD1998}, this corresponds to $A_K\sim0.5$ which is about a factor of 2 in flux.

\section{Analysis}\label{sec:results}
\subsection{Comparison to Previous Observations}

Figs.~\ref{fig:channel} and \ref{fig:hcochannel} show HCN and HCO$^+$ diffuse emission starting from $-20$ km s$^{-1}$ in the north and shifting southward with higher velocity before falling off near 120 km s$^{-1}$. CO channel maps show similar kinematic structure, albeit over a larger spatial region \citep{BIMASONG2,SBEL2006}.   
The HCN emission is largely confined to the innermost 20'', consistent with previous maps of this galaxy \citep{HB1997a}. Observations with a beamize of $\sim$1'' show that the HCN emission breaks down into a central clump elongated in the east-west direction, and a weaker emitting region 3'' east and 1'' south of the center \citep{SBED2007}.

The integrated intensities we measured  are broadly consistent with those compiled by \citet{M2006} convolved to a single dish 21'' beam pointed at the center of NGC 6946; he found  $I_{\rm HCN}=11$ K km s$^{-1}$, $I_{\rm HCO+}=13$ K km s$^{-1}$, and $I_{\rm CO}=220$ K km s$^{-1}$. At this resolution, we measured $I_{\rm HCN}=13$ K km s$^{-1}$, $I_{\rm HCO+}=14$ K km s$^{-1}$, and $I_{\rm CO}$= 280 K km s$^{-1}$. Differences between these two sets of observations can be explained by errors in flux calibration and single dish pointing. When measured with a single-dish telescope, a typical spiral  galaxy has an $I_{\rm HCN}/I_{\rm CO}$ ratio of 0.02--0.06 in its bulge \citep{HB1993,GS2004}.  The integrated intensities for NGC 6946 result in ratios that are well within this range.

\subsection{Ratio Maps}\label{sec:ratio}
\begin{figure}
\includegraphics[angle=90,scale=.4]{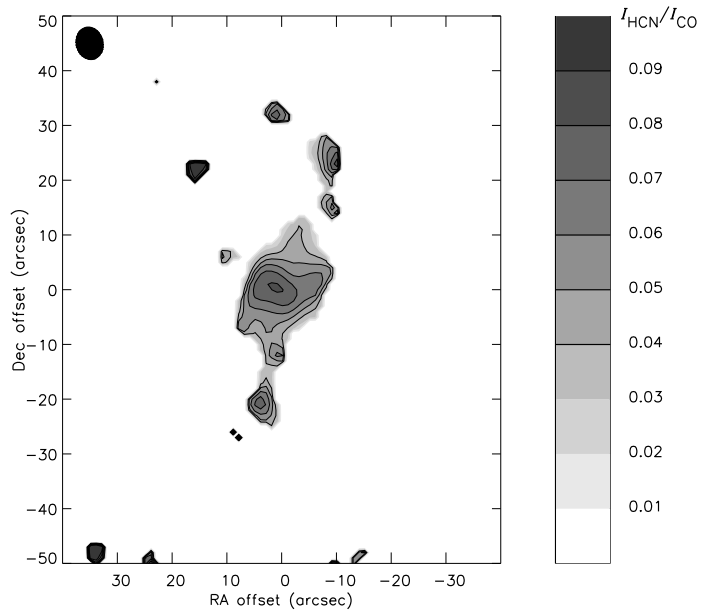}
\caption{\label{fig:ratiomap} Emission intensity ratio $I_{\rm HCN}/I_{\rm CO}$ on the sky coordinates. The contour levels are shown in the colorbar on the right, and the beam is shown in the upper left hand corner. For ease of reading, contours starting at 0.04 are also marked with a solid line.  The features seen at locations away from the central emission that are not seen in the HCN map in Fig.~\ref{fig:intmap} have a low value of $I_{\rm HCN}$.
}
\end{figure}

The $I_{\rm HCN}/I_{\rm CO}$ ratio is useful because it is a qualitative indicator of the density of the emitting gas.
In order to construct ratios of the HCN and CO maps, we convolved the HCN channel map to a 6.0'' by 4.9'' Gaussian beam to match the CO observations, and then created a mask based on the CO data as described in \S \ref{sec:intensity}; call this mask $C_{\rm CO}$. We recalculated the HCN, HCO$^+$, and CO integrated intensity maps using $C_{\rm CO}$ in eqn.~\ref{eqn:intens}; for the remainder of the paper, $I_X$ will refer to the integrated intensity calculate with this new $C_{\rm CO}$ mask.  We applied the same mask to all three molecules to avoid biasing the ratios.
The ratio $I_{\rm HCN}/I_{\rm CO}$ is shown in Fig.~\ref{fig:ratiomap}.
We also constructed independent noise maps, $Q_X$, for $I_{\rm CO}$, $I_{\rm HCN}$, and $I_{\rm HCO+}$ using the following formulation:
\begin{equation}
Q_X^2(\alpha,\delta)=\left(\sigma_{{\rm rms},X}~dv~\sqrt{\sum_v C_{\rm CO}}\right)^2+\left(u_XI_X\right)^2
\end{equation}
where $dv$ is the channel velocity width and $u_X$ is the uncertainty due to flux calibration for each species.

The ratio in Fig.~\ref{fig:ratiomap} is plotted only at points where $I_{\rm HCN}>2Q_{\rm HCN}$ and $I_{\rm CO}>5Q_{\rm CO}$.
The ratio in the central 20'' peaks at $0.081\pm0.021$ about 1'' east of the pointing center and has a relatively smooth falloff with increasing angular distance from the center; also, the three highest contours appear to be elongated along the east-west axis more than along the north-south one, as expected from the high resolution map \citep{SBED2007}. Extensions in the HCN distribution 25'' north and south of the pointing center are prominent, despite appearing weakly in the channel maps. These extensions have small pockets of $I_{\rm HCN}/I_{\rm CO}$ ratios that peak at $\sim$0.08, and they coincide with the molecular arms seen in previous CO maps \citep[i.e.][]{BIMASONG1}.
Small patches of HCN emission further from the center of the map have high ratios because of the low intensity of CO emission in those regions; they also have relatively little HCN emission.

\citet{HB1997a} measured the $I_{\rm HCN}/I_{\rm CO}$ ratio in NGC 6946 and found a central ratio of $\sim0.10$ using data with nearly identical spatial resolution but somewhat poorer sensitivity.   The structure of their ratio map is similar to ours, with the highest ratios along an east-west extended structure, but in their observations the structure is saddle shaped. It has a peak ratio of $0.19$ and maxima $\sim$5'' east and west of the center. These saddle peaks are not reproduced in our observations, but they were the most uncertain features of the earlier map.

\begin{figure}
\includegraphics[angle=90,scale=.4]{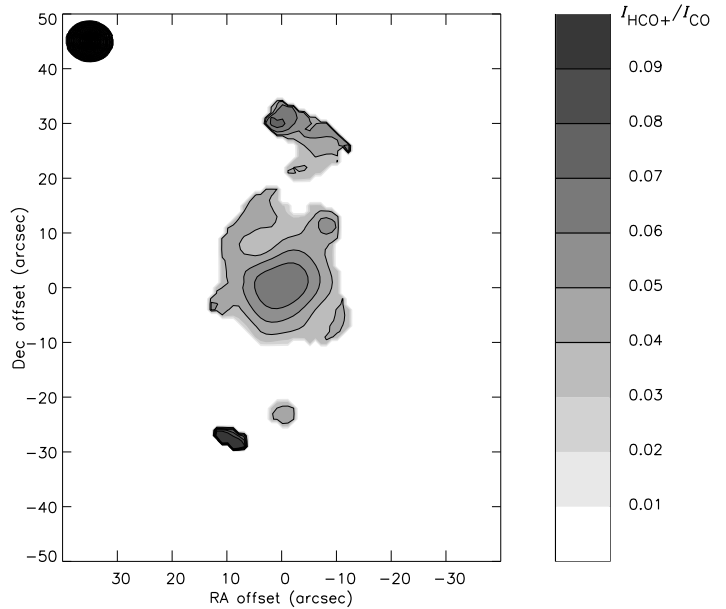}
\caption{\label{fig:hcoratiomap} Same as Fig.~\ref{fig:ratiomap} but for $I_{\rm HCO+}/I_{\rm CO}$.
}
\end{figure}

Using the same method, we also created a map of the $I_{\rm HCO+}/I_{\rm CO}$ ratio (Fig.~\ref{fig:hcoratiomap}). To make this map, we convolved the CO channel map to the larger beamsize of the HCO$^+$ observations.
The peak ratio in this map is $0.069\pm0.018$ measured less than an arcsecond from the pointing center. The higher peak ratio in the $I_{\rm HCN}/I_{\rm CO}$ map is due to its smaller beamsize  and is not physically significant. The  extensions to the north and south of the pointing center that appeared in the $I_{\rm HCN}/I_{\rm CO}$ map are present here as well.

\begin{figure}
\includegraphics[angle=90,scale=.4]{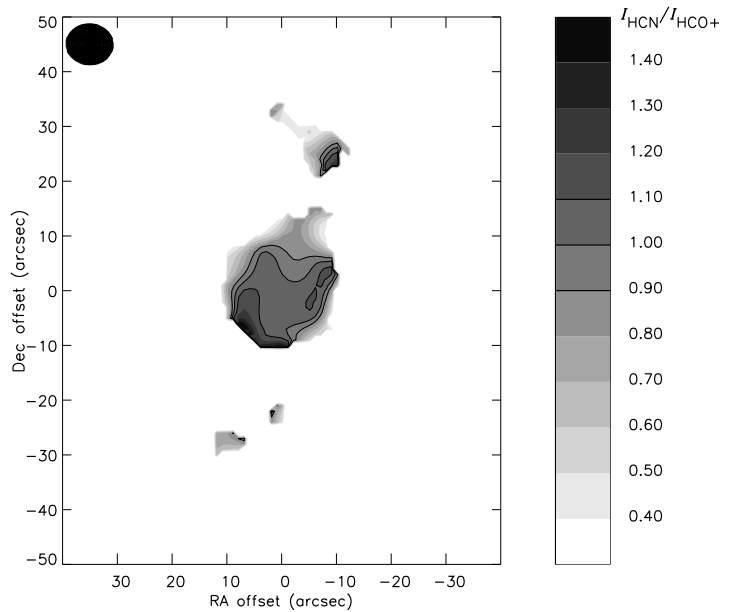}
\caption{\label{fig:hcnhcoratiomap} Same as Fig.~\ref{fig:ratiomap} but for $I_{\rm HCN}/I_{\rm HCO+}$. The 0.9, 1.0, and 1.1 contours  are marked with solid lines. }
\end{figure}

To compare the distribution of $I_{\rm HCN}$ and $I_{\rm HCO+}$, we constructed a final ratio map of the emission from the two dense gas tracers by convolving the HCN map to the HCO$^+$ beamsize (Fig.~\ref{fig:hcnhcoratiomap}). The ratio is only plotted at points where  $I_{\rm HCN}>2Q_{\rm HCN}$ and  $I_{\rm HCO+}>2Q_{\rm HCO+}$. Overall, the $I_{\rm HCN}/I_{\rm HCO+}$ ratio is close to 1 over the central 20''; since the two molecules have similar critical densities this was expected. In the cental regions there is a weak trend toward higher ratios in the southeast and lower ones in the northwest. In the extensions, the ratio appears to fall as low as 0.4, but the relatively low intensity of the emission from these features leads to large uncertainty in the ratio. 

\subsection{$I_{\rm HCN}$/$I_{\rm CO}$, $I_{\rm HCO+}$/$I_{\rm CO}$, and $I_{\rm HCN}$/$I_{\rm HCO+}$ Radial Falloffs}\label{sec:radius}
Although the $I_{\rm HCN}/I_{\rm CO}$ ratio map is roughly a map of the density of the emitting gas,
nearby points in the  map are not independent; their degree of correlation is quantified by the synthesized beam. To answer the question of how this qualitative measure of the density varies with radius, we constructed a map with uncorrelated data points.
We began by bilinearly interpolating the values of $I_{\rm HCN}$, $I_{\rm CO}$, $Q_{\rm HCN}$ and $Q_{\rm CO}$  on a hexagonal grid with 6.0''$\times$4.9'' spacing. This resampling results in intensities and errors that are only weakly correlated, since they are separated by a full beamwidth in the CO map, which has the larger beamsize. Throughout this paper, we will treat the points on this grid as uncorrelated.

In order to find how the ratio changes with distance from the galactic center, we corrected for the inclination of the galaxy, $i$. \cite{BIMASONG1} found
$i=54^\circ$ and a position angle of 65$^\circ$; \citet{T1988} found that the distance to NGC 6946 is 5.5 Mpc. Using these parameters, we  deprojected the coordinates on the sky to physical coordinates centered on the pointing center of the  observations. The inclination and position angle of NGC 6946 are a subject of disagreement, as  \citet{CA1988} found that NGC 6946 has $i=34\pm8^\circ$ and a position angle of 69$\pm$15$^\circ$. We  deprojected our data with both sets of parameters, and found that our results do not strongly depend on the assumed orientation of the galaxy.

\begin{figure}
\includegraphics[angle=0,scale=.43]{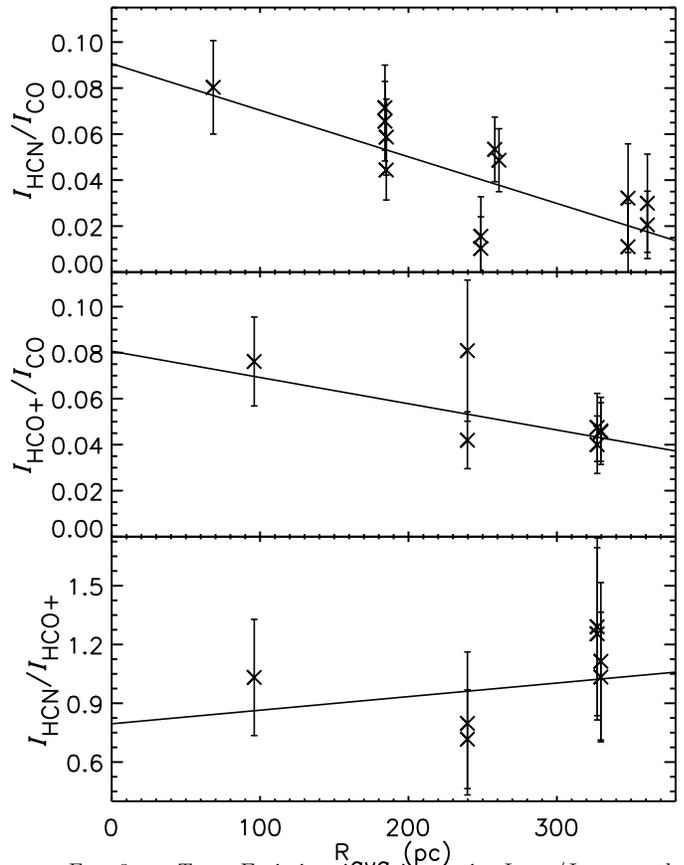}
\caption{\label{fig:ratio} {\it Top}: Emission intensity ratio $I_{\rm HCN}/I_{\rm CO}$ on the 6.0''$\times$4.9'' hexagonal grid. The radial coordinate is the deprojected physical radius averaged over a circular beam centered at each gridpoint. The units of the ratio are K km s$^{-1}$/K km s$^{-1}$. The line is the fit from eqn.~\ref{eqn:hcnfit}. {\it Middle}: The same plot for $I_{\rm HCO+}/I_{\rm CO}$ on the 8.6''$\times$7.5'' grid. The line is the fit from eqn.~\ref{eqn:hcofit}. {\it Bottom}: The same plot for $I_{\rm HCN}/I_{\rm HCO+}$. The line is the fit from eqn.~\ref{eqn:hcnhcofit}.}
\end{figure}

Fig.~\ref{fig:ratio} ({\it Top}) plots the $I_{\rm HCN}/I_{\rm CO}$ ratio as a function of average physical distance from the pointing center for the innermost points in the independently spaced hexagonal grid. The average physical distance, $R_{\rm avg}$ is determined for each point by averaging the radius over an ellipse with major and minor axes equal to the  grid spacing; we did this primarily to get an accurate average distance for the central point.  For the innermost points the error is dominated by the uncertainty in the flux calibration,  while at larger radii the channel noise is dominant. The ratio falls off strongly as a function of $R_{\rm avg}$. Ratios at radii larger than 380 pc suffer from systematic errors, and are not included in this analysis. If we sum the integrated intensities of all the points with $R_{\rm avg}$ between 380 and 500 pc, we find a ratio of $0.018\pm0.006$. This ratio is comparable to that found at the outermost points in Fig.~\ref{fig:ratio} ({\it Top}), which suggests that the ratio flattens out at larger $R_{\rm avg}$.
However, non-axisymmetric features with elevated ratios such as the molecular arms are a more likely reason for this trend.

We performed a linear $\chi^2$ fit on the $I_{\rm HCN}/I_{\rm CO}$ ratio as a function of $R_{\rm avg}$ for the inner 380 pc, and found:
\begin{equation}\label{eqn:hcnfit}
\frac{I_{\rm HCN}}{I_{\rm CO}}=0.091\pm0.016-(2.0\pm0.6)\times10^{-4}~R_{\rm avg}.
\end{equation}
The slope of this ratio is different from zero with 3$\sigma$ significance.
Note that by reducing the ratio to a function of $R_{\rm avg}$, we ignored any differences in physical conditions between points in the bar and those outside of it. We used a linear model because of its simplicity, but some other function likely fits the data better.

As noted in \citet{HB1997a}, the central $I_{\rm HCN}/I_{\rm CO}$ ratio in NGC 6946 is comparable to that measured in the inner 630 pc of the Milky Way, $0.081\pm0.004$ \citep{JHPB1996}. This suggests that the physical conditions, and therefore the molecular gas distribution, in the two systems are similar. 

To make the same plot for the $I_{\rm HCO+}/I_{\rm CO}$, we  constructed a new hexagonal grid because
 the HCO$^+$ data is of a lower resolution than the other species. We followed the prescription described above and create a new grid  with 8.6''$\times$7.5'' spacing. The falloff in the $I_{\rm HCO+}/I_{\rm CO}$ ratio with radius on this grid is shown in Fig.~\ref{fig:ratio} ({\it Middle}). The fit to this ratio is:
\begin{equation}\label{eqn:hcofit}
\frac{I_{\rm HCO+}}{I_{\rm CO}}=0.081\pm0.015-(1.1\pm0.5)\times10^{-4}~R_{\rm avg}.
\end{equation}
The slope of this ratio is different from zero with 2$\sigma$ significance. 

The strong falloff in the $I_{\rm HCN}/I_{\rm CO}$ and $I_{\rm HCO+}/I_{\rm CO}$ ratios suggest conditions in the inner 100 pc are quite different from those further from the center, and markedly different from those in the disk. Thus, there is no reason to expect that the molecular gas in the nucleus is distributed similarly to that in the disk.

Finally, on the coarse grid, we plot the ratio $I_{\rm HCN}/I_{HCO+}$ (Fig \ref{fig:ratio} ({\it Bottom})). The fit to these points is given by:
\begin{equation}\label{eqn:hcnhcofit}
\frac{I_{\rm HCN}}{I_{\rm HCO+}}=0.80\pm0.26+(6.9\pm10.0)\times10^{-4}~R_{\rm avg}.
\end{equation}
Note the large uncertainty in the slope; it is consistent with zero. We performed this last fit mainly as a consistency check; since HCN and HCO$^+$ are both dense gas tracers with similar critical densities, their ratio should be roughly constant with $R_{\rm avg}$. This is indeed what we observe.

\subsection{Molecular Line Ratios}

Since emission from CO, HCN, and HCO$^+$ all trace the molecular gas, there should be correlations between their integrated intensities. Following \citet{GS2004a}, we fit the relationships between the integrated intensities of the three dense gas tracers. We cannot use the linear $\chi^2$ fitting routine from the previous section because the points have errors in all parameters; furthermore, power laws are straightforward to fit in log-log space, but there the errors are asymmetric. For these reasons, we fit the relationships between the integrated intensities using a Monte Carlo fitting routine. A position for each point was randomly chosen using the calculated errors, and then a linear least squares fit was performed to solve for the fit parameters. For each fit, we repeated this process 10$^7$ times and took the mean and standard deviation of the fitted parameters.
 
\begin{figure}
\includegraphics[angle=0,scale=.43]{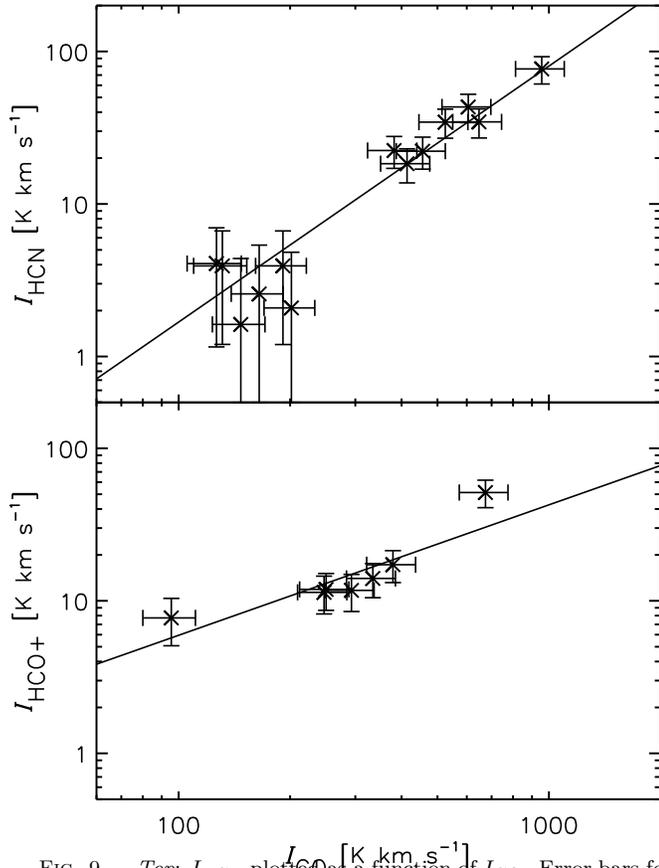}
\caption{\label{fig:covhcn} {\it Top}: $I_{\rm HCN}$ plotted as a function of $I_{\rm CO}$. Error bars for both molecules include random noise and the error in flux calibration. We plot the fit from eqn.~\ref{eqn:covhcnfit} as a solid line. {\it Bottom}: $I_{\rm HCO+}$ plotted as a function of $I_{\rm CO}$. Error bars for both molecules include random noise and the error in flux calibration. We plot the fit from eqn.~\ref{eqn:covhcofit} as a solid line. }
\end{figure}

 We begin with the relationship between $I_{\rm HCN}$ and $I_{\rm CO}$ (Fig.~\ref{fig:covhcn} ({\it Top})): 
\begin{equation}\label{eqn:covhcnfit}
\log_{10}{I_{\rm HCN}}= (1.7\pm0.3)\log_{10}{I_{\rm CO}}-3.1\pm0.8.
\end{equation}
Points in the lower integrated intensity regions have comparably larger errors in $I_{\rm HCN}$, so the poor fit in that region is of little significance. The overall fit suggests a power law relation with a slope significantly steeper than linear.

The same plot for HCO$^+$ is shown in Fig.~\ref{fig:covhcn} ({\it Bottom}). The fit for this relation is given by:
\begin{equation}\label{eqn:covhcofit}
\log_{10}{I_{\rm HCO+}}= (0.8\pm0.2)\log_{10}{I_{\rm CO}}-0.9\pm0.6.
\end{equation}
The single power law appears to fit less well here than in Fig.~\ref{fig:covhcn} ({\it Top}), but there are not enough independent data points to try a more complicated function.

These fits are an alternate method of presenting the variation in the $I_{\rm HCN}/I_{\rm CO}$ and $I_{\rm HCO+}/I_{\rm CO}$ ratios. In other words, the value of the slope in eqn.~\ref{eqn:covhcnfit} was not a surprise; it was clear from the shape of the ratio map center (Fig.~\ref{fig:ratiomap}) that the slope would be steeper than linear. We expected a similar slope in the HCO$^+$ fit, but the best fit to the data was a near-linear power law.  Since this fit is on the coarser grid, there are fewer points to constrain the relation, and the effect of fluctuations in the integrated intensities is greater. We suspect this results in an incorrect slope. The relatively poor quality of the fit in Fig.~\ref{fig:covhcn} ({\it Bottom}) lends credence to this argument.

Physically, what does this analysis of the molecular line integrated intensities and ratios imply about the distribution of molecular gas? A steeper than linear correlation between a dense gas tracer and CO suggests that  the proportion of dense gas  rises as $I_{\rm CO}$ increases, and also as distance from the center decreases. To understand what could cause this, we now investigate the physical conditions in the emitting region.

\subsection{Hydrostatic Midplane Pressure}
We constructed a measure of the hydrostatic midplane pressure to compare with the dense gas tracer emission.
We first  convolved the corrected $K_S$ image to the same beam as the CO observations, and sampled the resulting image on the hexagonal  grid. The stellar surface density at each point is then given by:
\begin{equation}
\log_{10}\frac{\Sigma_\star}{M_\odot~ \rm{pc}^{-2}}=-0.4\mu_{K_S}+9.62+\log_{10}\cos i
\end{equation}
where $\mu_{K_S}$ is the magnitude of flux in the $K_S$ band, in magnitudes per square arcsecond (BR06).

BR06 used the following formula for the hydrostatic pressure, $P_{\rm hyd}$, in the midplane of a disk:
\begin{eqnarray}\label{eqn:phyd}
\frac{P_{\rm hyd}}{k}&=&272~{\rm cm}^{-3}~{\rm K} \left(\frac{\Sigma_g}{M_\odot~{\rm pc}^{-2}}\right)
\left(\frac{\Sigma_\star}{M_\odot~{\rm pc}^{-2}}\right)^{0.5}\nonumber\\
&&\times\left(\frac{v_g}{{\rm km~s}^{-1}}\right)\left(\frac{h_\star}{\rm pc}\right)^{-0.5}.
\end{eqnarray}
The calculation of the gas and stellar surface densities, $\Sigma_g$ and $\Sigma_\star$ respectively, is described in the same paper. 
Eqn.~\ref{eqn:phyd} provides a rough estimate of the hydrostatic midplane pressure, since it was derived for a disk geometry and we used it on an elliptical stellar distribution.  
We used the CO integrated intensity to determine the molecular gas content, despite the possibility that using a constant X-factor with CO will overstate the molecular gas content in high density regions \citep{R2000,BIMASONG1}. We did this because we need a measure of the pressure that is independent of  $I_{\rm HCN}$ and $I_{\rm HCO+}$. The molecular gas surface density is larger than 50 $M_\odot$ pc$^{-2}$ throughout the inner 50'', and within 10'' of the pointing center it is larger than 500 $M_\odot$ pc$^{-2}$
We used a constant central HI gas surface density (corrected for inclination) of 3.9 $M_\odot$ pc$^{-2}$ \citep{TY1986}, which is much smaller than the molecular gas durface density. The gas velocity dispersion, $v_g$ is seen in face-on galaxies to be roughly constant \citep{SV1984,DHH1990,PR2007}; here we assume $v_g$ is 8 km~s$^{-1}$ throughout NGC 6946.
Note that this is generally interpreted as a large-scale turbulent velocity width, which is different from the velocity width within an individual cloud. This distinction will be important below.
 \citet{RV1995} deconvolved a K'-band image of the center of NGC 6946 into components, and found evidence for a 3.5'' oval distortion inside a 15'' bulge.
 We used the size of this smaller component to estimate the stellar scale height, $h_\star$. 

\begin{figure}
\includegraphics[angle=90,scale=.5]{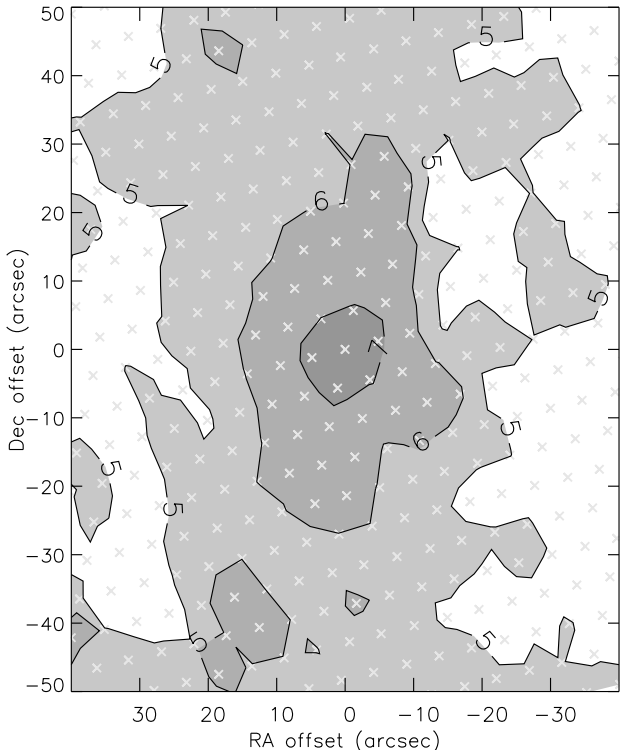}
\caption{\label{fig:pressure} A plot of $\log_{10}(P_{\rm hyd}/k)$ as a function of position on the 6''$\times$4.9'' hexagonal grid. The units of $P_{\rm hyd}/k$ are cm$^{-3}$ K. The  grid points are overlaid in light grey x's.
}
\end{figure}

The hydrostatic midplane pressure is the pressure required to balance the gravitational force of the stars; this pressure generally comes from turbulent or thermal motion of the gas. Note that $P_{\rm hyd}$ is the minimum required pressure, and the actual pressure in the midplane could be higher. Fig.~\ref{fig:pressure} shows the hydrostatic midplane pressure calculated as a function of position on the sky;
it peaks at $\sim5\times10^7$ cm$^{-3}$ K and falls off by more than an order of magnitude in the inner 10''. The uncertainty in the $K_S$ band flux due to dust discussed  in \S \ref{sec:2mass} corresponds to a small uncertainty in the pressure because the stellar surface density enters with a square root.

\subsection{Number Densities from Photon Dominated Region Models}\label{sec:PDR}

We used a code that relies on the $I_{\rm HCN}/I_{\rm HCO+}$ line ratio as sampled on the 8.6''$\times$7.5'' grid to calculate a number density of molecular hydrogen. This number density, $n_{\rm line}$, is specifically indicative of the physical conditions in the dense gas tracer emitting region, since it depends only on emission from tracers with high critical densities. 

We utilized a Photon Dominated Region (PDR)
code to calculate the thermal balance (with line transfer)
self-consistently with the chemical balance through iteration \citep{MS2005}. Both the density and ultraviolet radiation field $G_0$ were varied over a large
grid of models \citep{MSI2007}. 
Using the chemical and thermal structure of the PDR models along with a radiation transfer code \citep{PS2005,PS2006}, \citet{MSI2007} calculated the line intensities and ratios for
a number of species including HCN and HCO$^+$. We
compared these ratios to those observed for the center of NGC 6946 to find $n_{\rm line}$.

We found points in the inner 380 pc  have number densities $5\times10^4~{\rm cm}^{-3}~\lesssim n_{\rm line}\lesssim 10^5$ cm$^{-3}$ and average cloud temperatures $T\sim20$ K for $G_0=100$.
Choosing larger values of $G_0$ will raise $T$; for example, if $G_0=1000$, $T\sim45$ K. The value of $G_0$ does not affect $n_{\rm line}$.
\begin{figure}
\includegraphics[angle=90,scale=.32]{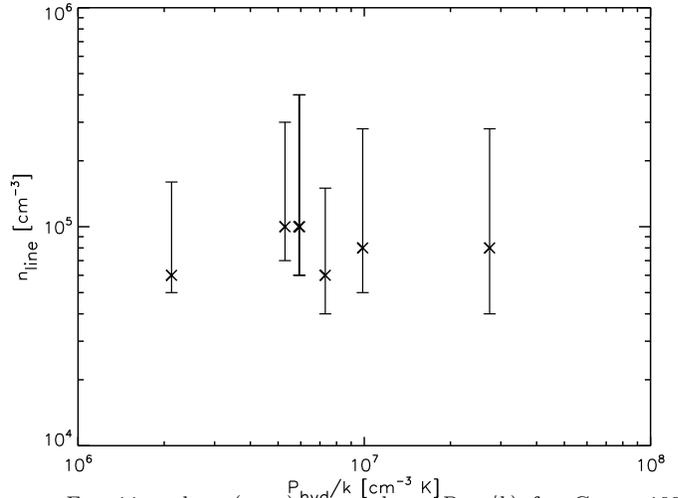}
\caption{\label{fig:pvp} $\log_{10}(n_{\rm line})$ versus  $\log_{10}(P_{\rm hyd}/k)$ for $G_0=100$. We calculated the error bars by running the 1$\sigma$ limits of the $I_{\rm HCN}/I_{\rm HCO+}$ ratio through the PDR code. 
}
\end{figure} 
Note that \citet{SGG1991} constrained the radiation field for the central 55'' of NGC 6946, and found $G_0\sim500$, consistent with our assumptions. We plot $n_{\rm line}$ as a function of $P_{\rm hyd}$ in Fig.~\ref{fig:pvp}.
The number density of the dense gas does not appear to vary as a function of $P_{\rm hyd}$ or $R_{\rm avg}$. We explore the implications of this below.

 \section{Discussion}\label{sec:discuss}
We discuss two plausible hypotheses for the distribution of dense gas in the central kpc of NGC 6946. The dense gas could be concentrated in self-gravitating clumps interspersed throughout a less dense molecular gas medium. Alternatively, the HCN and HCO$^+$ emission could come from a homogeneous component bound not by self-gravity, but by the ambient pressure.

\subsection{Self-gravitating Dense Gas and Implications for Star Formation}

One hypothesis consistent with our observations is that the  HCN and HCO$^+$ emission comes from self-gravitating dense clumps interspersed throughout less dense molecular gas traced by CO. 
The internal turbulent pressure of these density peaks is given by:
\begin{equation}
P_{\rm turb}\approx\rho v_{\rm int}^2
\end{equation}
where $v_{\rm int}^2$ is the internal gas velocity dispersion and $\rho$ is the density of the gas. In these peaks, $P_{\rm turb}$ is
much higher than the ambient pressure as measured by $P_{\rm hyd}$, so without self-gravity the density peaks would be unstable. This structure is similar in distribution to  GMCs in the galactic disk, 
where HCN and HCO$^+$ are excited in small density peaks inside of a envelope traced by CO. We have assumed that the turbulent pressure dominates the thermal pressure in these peaks, as it does in the star forming cores of GMCs.

We used observations of the nearby Galactic PDR in Orion to estimate the Gaussian velocity width of the peaks, $v_{\rm int}=2.7$ km s$^{-1}$ \citep{B1978a,TH1985}. In \S \ref{sec:PDR} we found that the number density in the inner region of NGC 6946 is constant with $P_{\rm hyd}$ and $R_{\rm avg}$; using $n_{\rm line}\sim10^5$ cm$^{-3}$ implies $P_{\rm turb}\sim10^8$ cm$^{-3}$ K,  larger than the calculated hydrostatic midplane pressure everywhere in the galaxy.
Therefore, self-gravity likely plays an important role in holding the density peaks together. 

 If the density peaks are self-gravitating, we can equate the force per unit volume due to turbulent pressure and gravity:
 \begin{equation}
 \frac{P_{\rm turb}}{r}\approx\frac{GM\rho}{r^2},
 \end{equation}
  where $M$ denotes the mass of the density peak and $r$ is the size of the peak.
 This reduces to the proportionality $M\propto v_{\rm int}^2 r$. Using the relation
 \begin{equation}
  n_{\rm line} \propto M/r^3 \propto{\rm constant}, 
\end{equation}
  we conclude that
  \begin{equation} 
  v_{\rm int}\propto r.
  \end{equation}
   Similarly, \citet{RB2005} found for giant molecular clouds in the center of M64 that $v_{\rm int}\propto r^{1.0\pm0.3}$, significantly different from GMCs in the disks of local group galaxies. M64 is a galaxy with high ambient pressure and molecular gas content  similar to the bulges of normal spirals. 
  The constraint was determined from observations of $^{13}$CO, which is better than $^{12}$CO at mapping dense gas due to optical depth effects, but does not map the density peaks as well as either HCN or HCO$^+$.  The authors then showed that the clouds were overpressurized compared to the hydrostatic midplane pressure by calculating the internal pressure from their measurements and assuming turbulent pressure was dominant. Here we find the density peaks are apparently overpressurized and that their number densities are constant, and follow the argument in the opposite direction to derive the power law they constrained experimentally. 

In the Milky Way, \citet{SRBY1987} used CO observations of disk GMCs to find that the surface density of the cloud is approximately constant with cloud size, and therefore the average density is inversely related to the cloud size. They constrained the relation between the size of the cloud, $s$, and its velocity dispersion to be $v_{\rm int}\propto s^{0.5\pm0.05}$ (note the distinction between size of the cloud $s$ and size of the density peak $r$). Since CO traces gas less dense than HCN and HCO$^+$, these results are not necessarily in conflict; density peaks in Milky Way clouds may follow the same relation as the density peaks in NGC 6946 even though the clouds as a whole do not.

As $R_{\rm avg}$ decreases,
the rise in both $I_{\rm HCN}$ and $I_{\rm HCO+}$ implies that the amount of dense gas 
 available for star formation is increasing. Since the density of the peaks is constant, this could result from an increase in  the size of individual density peaks,  in the number of density peaks above the necessary threshold, or some combination thereof.  
If the effect is due to the  number of density peaks increasing, this may imply a fundamental unit of star formation, as discussed in \citet{GS2004a} and \citet{WEGSS2005}. However, \citet{KT2007} pointed out a potential hole in this argument: the emission of any molecule with a critical density higher than the median density is expected to be correlated with star formation, regardless of whether the emission and star forming regions are directly associated. Further investigation would be aided by comparing our integrated intensities with the FIR intensity at each point. Unfortunately, FIR observations with 6'' resolution are not yet possible, even with the Spitzer space telescope.

\subsection{Homogeneous and Pressure Bound Dense Gas Tracers}
Alternatively, the emission from HCN and HCO$^+$ could come from homogeneous gas bound by the ambient pressure. This situation is similar to what is observed in the Milky Way center, where we see emission from HCN and CS everywhere we see CO, within the sensitivity limits \citep[and references therein]{HB1997b}. That is, despite the best efforts to resolve the dense gas tracer emission into  clumps, it still appears to be diffusely distributed. The mean number density of the gas appears to be larger than the critical density of the dense gas tracers, and there is no need to appeal to self-gravitating dense clumps to explain the observed emission. In this model, the dense gas filling factor is much larger than that of the model described in the previous section.

 To construct a similar situation in the central kpc of NGC 6946, support of the molecular gas layer against the gravitational potential must come not from turbulent pressure, but from thermal pressure:
 \begin{equation}
P_{\rm therm}=nkT.
\end{equation}
    The central hydrostatic midplane pressure is $5\times10^7$ cm$^{-3}$ K; if the support comes from turbulence with a characteristic speed of 3 km s$^{-1}$, the implied density would be $\sim2\times10^4$ cm$^{-3}$, which is too low to excite HCN. 
 Instead, if the thermal pressure provides the support, then for kinetic temperatures in the range of 50--200 K, consistent with those in the center of the Milky Way, the derived densities are $2.5\times10^5-1\times10^6$ cm$^{-3}$. These densities are high enough to excite the dense gas tracers, and are marginally consistent with the number densities calculated via the PDR code.  Temperatures as low as 50 K are consistent with the PDR calculations with $G_0$ $\sim$ 1000. Higher temperatures would require larger values of $G_0$, but the radiation field is not constrained by our observations.

Even at a distance of $R_{\rm avg}\sim400$ pc from the center, thermal pressure is likely to play an important role. In this region, we know at least some of the gas has $n\sim10^5$ cm$^{-3}$, because we observe emission from the dense gas tracers. Using the PDR code tells us that the temperature of this dense gas is $\sim30$ K, which implies a thermal pressure of $3\times10^6$ cm$^{-3}$ K. The hydrostatic midplane pressure for these points is $\lesssim 5\times10^6$ cm$^{-3}$ K, so thermal pressure in this region can provide a significant portion of the necessary hydrostatic support, if not all of it.

Determining whether hydrostatic support at the galactic center comes from turbulent or thermal pressure would point out which of the self-gravitating dense gas model or the homogeneous and pressure bound dense gas tracer model is correct. We can do this by constraining $G_0$ for the central 10'' of NGC 6946 (see \citet{KMB2005} for details on constraining $G_0$); if $G_0$ is large ($\sim10^4$) the gas is likely warm and thermal pressure will dominate, and if it is small ($\sim10^2$) the gas is cool and turbulent pressure is more important. Unfortunately, the constraint $G_0\sim500$ from \citet{SGG1991} was calculated from C {\footnotesize II} observations with a resolution of 55''; better resolution than this is necessary to differentiate between the two models. 

\subsection{Hydrostatic Midplane Pressure and Line Intensities}

\begin{figure}
\includegraphics[angle=0,scale=.43]{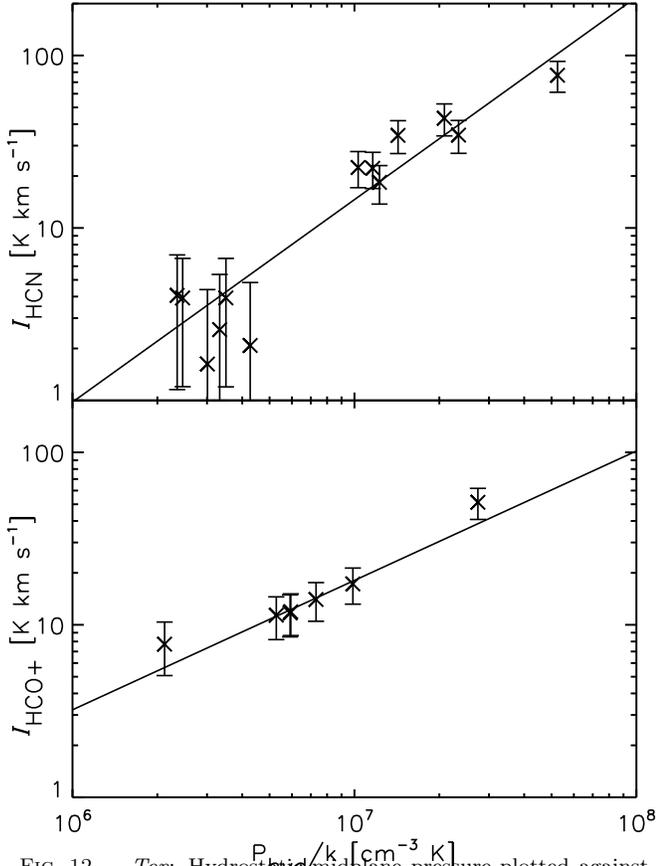}
\caption{\label{fig:pvhcnhco} {\it Top}: Hydrostatic midplane pressure plotted against $I_{\rm HCN}$. 
 The solid line is the fit described in eqn.~\ref{eqn:pvhcnfit}. {\it Bottom}: Hydrostatic midplane pressure plotted against $I_{\rm HCO+}$. The solid line is the fit described in eqn.~\ref{eqn:pvhcofit}}
\end{figure}

There is a robust correlation between  $I_{\rm HCN}$ and $P_{\rm hyd}$ (Fig.~\ref{fig:pvhcnhco} ({\it Top})). We fit the relation between the two variables with the Monte Carlo routine, which yielded:
\begin{equation}\label{eqn:pvhcnfit}
\log_{10}I_{\rm HCN}=-7.0\pm1.3+(1.2\pm0.2)\log_{10}\left(\frac{P_{\rm hyd}}{k}\right),
\end{equation}
where $P_{\rm hyd}$ is  in cm$^{-3}$ K.
It is clear that the total emission from HCN increases as $P_{\rm hyd}$ rises.

We performed the same fit for $I_{\rm HCO+}$ on the coarser 8.6''$\times$7.5'' grid (Fig.~\ref{fig:pvhcnhco} ({\it Bottom})). The Monte Carlo fit yielded:
\begin{equation}\label{eqn:pvhcofit}
\log_{10}I_{\rm HCO+}=-4.0\pm1.0+(0.7\pm0.1)\log_{10}\left(\frac{P_{\rm hyd}}{k}\right),
\end{equation}
where $P_{\rm hyd}$ is  in cm$^{-3}$ K.
Again, 
it is clear that $I_{\rm HCO+}$ rises with increasing $P_{\rm hyd}$. These two relations appear to have significantly different slopes, but that may be due to the differing resolution of the $I_{\rm HCN}$ and $I_{\rm HCO+}$ maps.

These fits, along with the radial fits in \S \ref{sec:radius} and the hydrostatic midplane pressure map (Fig.~\ref{fig:pressure}), indicate that the molecular gas in the central kiloparsec of NGC 6946 is under high pressure, confirming a prediction made in \citet{SB1992}. Emission from dense gas tracers like HCN and HCO$^+$ is widespread, and their integrated intensity ratios to CO are larger than those further from the center. 
This gas is in a markedly different state than the local interstellar medium in the Galactic disk where the pressure is approximately three orders of magnitude lower. The structure of the molecular gas distribution, whether it is self-gravitating or homogeneously distributed and  pressure bound,   also bears little resemblance to the local interstellar medium.

\subsection{Hydrostatic Midplane Pressure and Molecular Gas}

\begin{figure}
\includegraphics[angle=90,scale=.32]{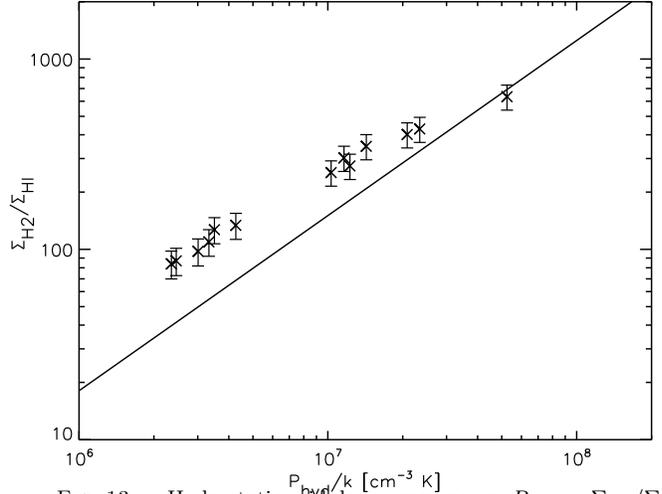}
\caption{\label{fig:pvratio} Hydrostatic midplane pressure vs. $R_{\rm mol}=\Sigma_{\rm H_2}/\Sigma_{\rm HI}$. The HI surface density has been assumed constant over the range of this plot. The error bars are determined only from the error in $\Sigma_{\rm H_2}$; we do not include error in the HI surface density. The line is the fitted relation for normal galaxies from BR06.}
\end{figure}

The hydrostatic midplane pressure has a nearly linear relationship with the $\Sigma_{\rm H2}/\Sigma_{\rm HI}$ ratio, $R_{\rm mol}$, over three orders of magnitude in pressure \citep{BR2004,BR2006}. Fig.~\ref{fig:pvratio} plots $R_{\rm mol}$ against the hydrostatic midplane pressure for the points in the inner central 350 pc of NGC 6946, along with the fit to normal galaxies from BR06. The fit follows the points reasonably well, though the data are slightly displaced toward higher $P_{\rm hyd}$. However, BR06 did see comparable scatter in their original sample around this relation. This paper supplements the previous work by adding more points at the high pressure (10$^6$-10$^8$ cm$^{-3}$ K) end of the relationship, albeit with the assumption that the HI surface density is constant. With such large molecular gas surface densities, the HI gas should not contribute a significant amount to the total gas surface density anyway.

Both $P_{\rm hyd}$ and $R_{\rm mol}$ depend on $I_{\rm CO}$, and plotting one as a function of the other can lead to misleading correlations.
However, the fit found in BR06 was determined from many points with $R_{\rm mol}\sim1$, and the authors showed that the relationship was physically significant in spite of the self-correlation issues. We include this plot to demonstrate the continued applicability of the fit in the high pressure regime.

\section{Conclusions}

Using the BIMA and CARMA interferometers, we  produced high-resolution maps of the dense gas tracers HCN and HCO$^+$ in NGC 6946. The maps are rich with detail, and we have 13 independent data points across the inner $\sim$750 pc of the galactic center in HCN and CO, and 7 in HCO$^+$. The integrated intensities of HCN and HCO$^+$ peak near the center of the galaxy, and fall off from there. 

Using BIMA SONG CO 1-0 and 2MASS $K_S$ band images, we calculated an approximation to the hydrostatic midplane pressure in the center of NGC 6946. 
We used a PDR model to calculate the number density in the dense gas regions traced by HCN and HCO$^+$, and found that the number density in these regions does not vary as a function of radius or the surrounding hydrostatic midplane pressure. We explored two hypotheses for the distribution of the dense gas in the central kpc of NGC 6946. If the dense gas is concentrated in clumps, we showed that self-gravity implies that $v_{\rm int}\propto r$. However, we also explored another plausible dense gas distribution; the dense gas could instead be homogeneously distributed throughout the galactic center if thermal pressure provides enough support for hydrostatic equilibrium. 
 We then demonstrated clear correlations between the HCN and HCO$^+$ integrated intensities and the midplane pressure.  Finally, we confirmed the validity of the relation between hydrostatic midplane pressure and the molecular to atomic gas ratio found in BR06 in the high pressure regime.

\acknowledgements
ESL, TTH,  and LB are supported by NSF grant AST-0540567. RM is supported by NSF grant AST-0507423. We would like to thank the staff at the BIMA and CARMA observatories for their help in making the HCN and HCO$^+$ observations. Thanks to Jin Koda, Erik Rosolowsky, Dick Plambeck, Alberto Bolatto, Conor Laver, Josh Peek, and Karin Sandstrom for helpful conversations.

\bibliographystyle{apj}
%\bibliography{/Users/elevine/Documents/work/papers/mybiblio}

\end{document}